\documentclass[fleqn,usenatbib]{mnras}

\usepackage[T1]{fontenc}
\usepackage{ae,aecompl}

\usepackage{graphicx}	% Including figure files
\usepackage{amsmath}	% Advanced maths commands
\usepackage{amssymb}	% Extra maths symbols

\title[Proto-magnetar jet engine for SNe Ic-bl]{Proto-magnetar jets as central engines for broad-lined type Ic supernovae}

\author[Shankar et al.]{
Swapnil Shankar,$^{1}$\thanks{E-mail: s.shankar@uva.nl}
Philipp M\"{o}sta,$^{1}$\thanks{E-mail: p.moesta@uva.nl}
Jennifer Barnes,$^{2}$ 
Paul C. Duffell$^{3}$ and
Daniel Kasen$^{4,5}$
\\
$^{1}$GRAPPA, Anton Pannekoek Institute for Astronomy and Institute of High-Energy Physics, University of Amsterdam, Science Park 904,\\
1098 XH Amsterdam, The Netherlands\\
$^{2}$Department of Physics, Columbia University, 538 W 120 Street, New York, NY 10025, USA\\
$^{3}$Department of Physics and Astronomy, Purdue University, 525 Northwestern Avenue, West Lafayette, IN 47907-2036, USA\\
$^{4}${Department of Physics and Department of Astronomy, University of California, Berkeley, 366 LeConte Hall, Berkeley, CA, 94720, USA}\\
$^{5}${Nuclear Science Division, Lawrence Berkeley National Laboratory, 1 Cyclotron Road, Berkeley, CA, 94720, USA}\\
}

\date{Accepted XXX. Received YYY; in original form ZZZ}

\pubyear{2021}

\begin{document}
\label{firstpage}
\pagerange{\pageref{firstpage}--\pageref{lastpage}}
\maketitle

\begin{abstract}
A subset of type Ic supernovae (SNe Ic), broad-lined SNe Ic (SNe Ic-bl), show unusually high kinetic energies ($\sim 10^{52}$~erg) which cannot be explained by the energy supplied by neutrinos alone. Many SNe Ic-bl have been observed in coincidence with long gamma-ray bursts (GRBs) which suggests a connection between SNe and GRBs. A small fraction of core-collapse supernovae (CCSNe) form a rapidly-rotating and strongly-magnetized protoneutron star (PNS), a proto-magnetar. Jets from such magnetars can provide the high kinetic energies observed in SNe Ic-bl and also provide the connection to GRBs. In this work we use the jetted outflow produced in a 3D CCSN simulation from a consistently formed proto-magnetar as the central engine for full-star explosion simulations. We extract a range of central engine parameters and find that the extracted engine energy is in the range of $6.231 \times 10^{51}-1.725 \times 10^{52}$~erg, the engine time-scale in the range of $0.479-1.159$~s and the engine half-opening angle in the range of $\sim 9-19\degr$. Using these as central engines, we perform 2D special-relativistic (SR) hydrodynamic (HD) and radiation transfer simulations to calculate the corresponding light curves and spectra. We find that these central engine parameters successfully produce SNe Ic-bl which demonstrates that jets from proto-magnetars can be viable engines for SNe Ic-bl. We also find that only the central engines with smaller opening angles ($\sim 10\degr$) form a GRB implying that GRB formation is likely associated with narrower jet outflows and Ic-bl's without GRBs may be associated with wider outflows. 
\end{abstract}


%%%%%%%%%%%%%%%%%%%%%%%%%%%%%%%%%%%%%%%%%%%%%%%%%%

%%%%%%%%%%%%%%%%% BODY OF PAPER %%%%%%%%%%%%%%%%%%

\section{Introduction}
Gamma ray bursts (GRBs) are short and intense flashes of gamma rays at cosmological distances \citep[e.g.][]{Fishman_1995}. They can be classified as short GRBs and long GRBs, depending on the duration of the burst. Core-collapse supernovae (CCSNe) are explosions of massive stars at the end of their lifetime, forming a neutron star or a black hole in the process \citep[e.g.][]{Woosley_2005}. The connection between SNe and GRBs has been theorized before \citep{Colgate_1968, Woosley_1993, Bohdan_1998}  but was only confirmed observationally with the discovery of SN 1998bw coincident with GRB 980425 \citep{Galama_1998}, which suggested a connection between the two phenomena. The SN-GRB connection has since become firmer with the nearly simultaneous discovery of SN2003dh with GRB 030329 \citep{Stanek_2003,Hjorth_2003,Matheson_2003}  and is now well established with additional observations, e.g. SN 2006aj/GRB 060218 \citep{Campana_2006,Modjaz_2006,Pian_2006,Sollerman_2006}, and SN 2010bh/GRB 100316D \citep{Chornock_2010,Starling_2011}. \\

All SNe that have been linked to GRBs belong to the  class of broad-lined Type Ic SNe \citep[SNe Ic-bl; e.g.][]{Woosley_2006,Modjaz_2011,Hjorth_2012,Cano_2017}. SNe Ic-bl have broad spectral lines indicating high photospheric velocities \citep[$\sim15,000 - 30,000$~km s$^{-1}$;][]{Modjaz_2016} and high kinetic energies \citep[$\sim 10^{52}$~erg; e.g.][]{Iwamoto_1998,Olivares_2012}. Their optical spectra show no H or He. The extreme kinetic energies involved in SNe Ic-bl challenge the underlying standard explosion mechanism, because the energy supplied by neutrinos is not sufficient to explain the high kinetic energies observed \citep[e.g.][]{Burrows_2021}. Jets from rapidly rotating protoneutron stars (PNS) formed in CCSNe explosions can provide the high kinetic energies observed in SNe Ic-bl and also provide the connection to GRBs \citep{Komissarov_2008,Metzger_2011,Wang_2016,Sobacchi_2017,Burrows_2021}. However, whether a single jet engine can explain both SNe Ic-bl and GRBs is still unclear. Many SNe Ic-bl have been observed without an accompanying GRB, which raises the question whether GRBs are present in all SNe Ic-bl. \cite{Modjaz_2016} found in a statistical study that SNe with an accompanying GRB have broader spectra compared to SNe without an observed GRB and that line of sight effects alone are not likely to explain the fraction of SNe Ic-bl with and without accompanying GRBs.\\

Some CCSNe explosions can lead to the formation of a rapidly-rotating PNS where vigorous convection coupled with rapid rotation forms very strong magnetic fields ($\sim 10^{15}$~G) due to magnetic field amplification via dynamo action \citep{Duncan_1992}. Simulations show that the spin-down of the rapidly-rotating PNS can supply energy to the jet outflow resulting in higher kinetic energies compared to a neutrino-driven SN explosion \citep{Mazzali_2014}. In principle, the SN explosion energy and light curves derived from CCSNe can be tested with self-consistent MHD simulations following the jet all the way to break-out of the stellar surface. However, this is currently numerically infeasible in multiple dimensions because of the difference in length scales of the PNS ($\sim 10$~km) and the progenitor star($\sim 10^{6}$~km), as well as the difference in time scales of jet formation ($\sim 0.1$~s) and jet breakout ($\sim 10$~s), which need to be resolved numerically leading to a spatial resolution of $\sim 0.1$~km and a corresponding time-step of $\sim 10^{-7}$~s. The current approach for full-star simulations is to excise some portion ($\sim 1000$~km) from the centre of the star and assume a hypothetical engine injecting energy into the rest of the star. This is known as the central engine paradigm \citep[e.g.][]{Suzuki_2019}.\\
           
Under the central engine paradigm, \cite{Barnes_2018} (\textit{B2018}, hereafter) combined  hydrodynamics and radiation-transfer simulations end-to-end to simulate a GRB jet driven SN Ic-bl. The success of this numerical setup depends on its ability to produce high kinetic energies and broad spectral features typical of SNe Ic-bl. For a presumed set of engine parameters, \textit{B2018} were successful in producing a SN Ic-bl that was roughly consistent with observations. In their work, they chose values for the central engine parameters consistent with observations. Whether such engine parameters are possible from PNS formation in CCSNe simulations remains to be investigated. We probe this in the current work. \\  
            
In this work, we use the data from a 3D magnetorotational CCSN simulation \citep{Moesta_2014} to estimate the engine parameters. We then use these parameters to perform hydrodynamic and radiation-transfer calculations. We closely follow the numerical setup of \textit{B2018} for the hydrodynamics and radiation transfer simulations. This is the first study to carry out an end-to-end CCSN simulation, hydrodynamics, and radiation transfer calculation in multiple dimensions. We find that the central engine parameters extracted from jet outflows of 3D CCSN simulation successfully produce a SN Ic-bl. This demonstrates that jets from PNS formation can be a viable engine for SN Ic-bl. \\
          
In section \ref{Numerical_Setup}, we describe the tools used in our numerical setup. In section \ref{Results}, we present the methodology and results for parameters extracted from the 3D CCSN simulation, as well as the results for SN observables. We discuss the obtained results in Section \ref{discussion}. 

\section{Numerical Setup}\label{Numerical_Setup}
We combine the results from a 3D magnetorotational CCSN simulation with a suite of advanced numerical codes to model a jet driven SN explosion and its emergent light curves \& spectra. We use the 3D CCSN simulation of \cite{Moesta_2014}  to estimate the engine parameters. We perform the hydrodynamic simulations with the 2D special relativistic {\scriptsize{JET}} code and carry out radiation transport with {\scriptsize{SEDONA}} to generate the light curves and spectra. {\scriptsize{JET}} takes as input the extracted engine parameters and gives as output the density, temperature and $^{56}$Ni mass distribution. We then use these as input to {\scriptsize{SEDONA}} to generate light curves and spectra. This numerical setup allows us to study a jet-driven SN explosion of the star, including the physics of core collapse, in multiple dimensions.
 
\subsubsection*{3D GRMHD CCSN simulation for central engine parameter estimation}  
\cite{Moesta_2014} perform three-dimensional general-relativistic magnetohydrodynamic (GRMHD) simulations of rapidly rotating strongly magnetized CCSNe. The simulation was performed in ideal GRMHD with the {\scriptsize{EINSTEIN TOOLKIT}} \citep{M,Loffler_2012}.   They employ a finite-temperature microphysical equation of state (EOS), using the $K_0=220$ MeV variant of EOS of \cite{Lattimer_1991}, and an approximate treatment of neutrino transport \citep{Connor_2010,Ott_2012}. They use the 25 $M_\odot$ presupernova model E25 \citep{Heger_2000} as the progenitor. They perform the simulations in full unconstrained 3D as well as those constrained to 2D, both of which start from identical initial conditions. They find that 2D and 3D simulations show fundamentally different evolutions. A strong jet-driven explosion is obtained in 2D. In contrast, the jet disrupts in full 3D and results instead in a broad lobar outflow. In this work, we use the results of the 3D simulation and estimate the central engine parameters from the lobar outflow.   

\subsubsection*{Hydrodynamics using {\scriptsize{JET}}}
{\scriptsize{JET}} \citep{Duffell_2013} is a variant of {\scriptsize{TESS}} \citep{Duffell_2011}, with a specific application to radial outflows. {\scriptsize{JET}} uses a mesh which moves outward radially, thus making it effectively Lagrangian in radial direction and able to accurately evolve flows over large dynamic length scales. This is very useful in the current work because we need to evolve the flow from $\sim 10^3$~km to $\sim 10^9$~km. We have used the most recent version of {\scriptsize{JET}} code for our hydrodynamics calculations.  Except for varying the central engine parameters, we keep other simulation parameters as in \textit{B2018}.\\

Radioactive decay of $^{56}$Ni is the source of luminosity for the SN, but {\scriptsize{JET}} does not include a nuclear reaction network to accurately model the synthesis of $^{56}$Ni during the hydrodynamic phase. \textit{B2018} provide a detailed description of $^{56}$Ni synthesis in the {\scriptsize{JET}} code, but we briefly reiterate it here because it is fundamental to the SN model. {\scriptsize{JET}} uses an approximate treatment to estimate $^{56}$Ni production using a simple temperature condition in which any zone where temperature exceeds a certain temperature, $T_{\text{max}}$, is assumed to burn to pure $^{56}$Ni. We use $T_{\text{max}} = 5 \times 10^9$~K as in \textit{B2018}.         

\subsubsection*{Radiation transport using {\scriptsize{SEDONA}}}
{\scriptsize{SEDONA}} is a 3D time dependent multi wavelength radiative transport code based on Monte Carlo techniques, which can be used to calculate SN observables from the hydrodynamic variables of a SN \citep{Kasen_2006}. {\scriptsize{SEDONA}} self-consistently solves the temperature structure of the ejecta and generates the temperature and composition dependent opacities required for photon transport. Our calculation assumes the ejecta is in local thermodynamic equilibrium (LTE). The code calculates the light curves \& spectra. It outputs the supernova's full spectral time-series, thus providing a link between the hydrodynamic calculation and SN observables. We have used the most recent version of {\scriptsize{SEDONA}} in a 2D axisymmetric setting. Our numerical setup for {\scriptsize{SEDONA}} calculations is the same as that of \textit{B2018}. 

\subsubsection*{Progenitor and Jet engine models}
We use the same progenitor and engine models as \textit{B2018}. A detailed description of the progenitor and engine models can be found there, but we briefly reiterate the relevant details here. The progenitor consists of an analytic model that reasonably approximates the major features of a stripped-envelope Wolf-Rayet star having zero-age main-sequence mass of 40 $M_{\odot}$ and solar metallicity. We excise the material interior to 0.015 $R_{\odot}$ of the star and set the density in the cavity to $10^{-3}$ times density at the cavity boundary. The density exterior to the cavity is given by: 

\begin{equation}
 \rho_{\text{init}}(r) = \frac{0.0615M_0}{R_0^3} (R_0/r)^{2.65}(1-r/R_0)^{3.5}
\end{equation}

where $R_0 = 1.6 R_{\odot}$ is the radius of the star and $M_0 = 2.4 M_{\odot}$ is the mass of the material outside the cavity. The composition of the progenitor in terms of mass fractions of various elements is shown in Table~\ref{tab:progenitor_composition}.\\

\begin{table}
	\centering
	\caption{Progenitor composition}
	\label{tab:progenitor_composition}
	\begin{tabular}{cccccc} 
		\hline
		 He   &   C   &   N  &   O   &   Ne    &   Mg \\ 
		 6.79e$-$3   &   2.27e$-$2   &   2.91e$-$5  &   9.05e$-$1   &   1.37e$-$2    &   8.46e$-$3 \\ 
		\hline
		   Si   &   S   &   Ar  &   Ca   &   Ti    &   Fe \\
		   2.69$-$2   &   1.04e$-$2   &   1.60e$-$3  &   6.63e$-$4   &   5.11e$-$7    &   3.50e$-$3 \\
		\hline	
	\end{tabular}
\end{table}

The engine is defined by the total energy injected, $E_{\text{eng}}$; the engine half-opening angle, $\theta_{\text{eng}}$; and the characteristic time-scale of the engine, $t_{\text{eng}}$. We taper off the engine exponentially as

\begin{equation}
L_{\text{eng}}(t) = \frac{E_{\text{eng}}}{t_{\text{eng}}} \times \exp[-t/t_{\text{eng}}]
\end{equation}

We estimate the values of $E_{\text{eng}}$, $\theta_{\text{eng}}$ and $t_{\text{eng}}$ using the 3D CCSN simulation from \cite{Moesta_2014}. The values of a few important {\scriptsize{JET}} and {\scriptsize{SEDONA}} parameters used in our numerical setup are listed in Table~\ref{tab:simulation_parameter_table}.

\begin{table*}
	%\centering
	\caption{A few important {\scriptsize{JET}} and {\scriptsize{SEDONA}} parameters }
	\label{tab:simulation_parameter_table}
	\begin{tabular}{llll} 
		\hline
		\multicolumn{2}{c}{{\scriptsize{JET}}} & \multicolumn{2}{c}{{\scriptsize{SEDONA}}} \\
		\hline
		 Parameter &  Value   & Parameter &  Value   \\
		\hline
		  Adiabatic Index & 4/3 &  Number of particles & $2 \times 10^{5}$ \\
		  Injected Lorentz Factor  & 50 &  Number of viewing bins & 9 (between 0 and $\pi$)  \\	
		  Energy-to-Mass Ratio  &  1000 &  Transport frequency grid (Logarithmic)$^a$  & \{$2.7\times 10^{13}$, $2\times 10^{16}$, 0.0006\} \\
		  Nozzle Size & $\sim 600$~km &  Spectrum frequency grid (Logarithmic)$^a$  & \{$3.0\times 10^{13}$, $2\times 10^{16}$, 0.005\} \\
		  Initial Number of Radial Bins & 512  & Start time & 0.5~d \\
		  Number of Angular Bins & 256 (between 0 and $\pi/2$)  & Stop time & 80.0~d \\
		   &   &    Maximum value of a timestep & 0.1~d\\
		   &   &    Number of velocity bins & 100 (between 0 and $0.2c$)\\
		\hline	
	\end{tabular}
	\\
	\footnotesize{$^a$ frequency grid = \{$\nu_{\text{start}}$(s$^{-1}$), $\nu_{\text{stop}}$(s$^{-1}$), $\Delta$\}, $d\nu=\nu\Delta$ }
\end{table*}

\subsubsection*{Reproduction of earlier results}
\textit{B2018} have already performed an end-to-end hydrodynamic and radiation-transfer simulation with a presumed set of engine parameters. In the current work, we add to this setup the engine parameters extracted from a 3D CCSN simulation, instead of  using a presumed set of parameters. We have used the most recent versions of {\scriptsize{JET}} and {\scriptsize{SEDONA}} for this work. To validate our methodology and isolate the effects of changes to {\scriptsize{JET}} and {\scriptsize{SEDONA}} on simulation outputs, we first reproduce the results of B2018 using our updated computational suite.  We find that our model spectra show no significant differences from the results of  \textit{B2018}. Our spectra have fairly broad lines representative of SNe Ic-bl, with enhanced broadening and blue-shifting for polar viewing angles at times less than $t_{\text{peak}}$. \\

\section{Results}\label{Results}
\subsection{Extraction of engine parameters}
Next, we extract from the CCSN simulation an effective engine time-scale, energy and half-opening angle for use in new simulations with {\scriptsize{JET}}. The PNS formed in the 3D CCSN simulation produces a wide-lobed outflow as the actual jet gets disrupted by an $m=1$ kink instability. It is this outflow which provides the energy for exploding the stellar material as a SN. We use this outflow to estimate the total energy, the half-opening angle and the characteristic time-scale of the central engine. Data for various physical parameters in the 3D CCSN simulation is available up to $\sim 130$~ms after bounce. However, this time-scale is much smaller than the time-scale involved in central engine and stellar SN dynamics. We therefore need to extrapolate the available data in order to get an estimate for the engine parameters. We use the spin-down rate of the PNS to estimate the time-scale of the central engine and we assume that the spin-down rate remains constant for $\sim1$~s after data availability. In reality, the accretion rate may deviate from the extrapolated behaviour at late times ($t\gtrsim 0.5$~s) \citep[for example see][]{Burrows_2020}. This could lead to a different engine behaviour at late times, which could lead to a different $t_{\text{eng}}$.

\subsubsection*{Estimation of total energy and characteristic time-scale}
The jet produced in the 3D CCSN simulation gets disrupted and a wide-lobed outflow of material forms instead. We need to estimate the central engine total energy from this resultant flow pattern. For that we identify the material that is gravitationally unbound from the newly formed PNS and can be considered ejected from it. It is this unbound material that provides the energy required for the explosion of the star. We define a fluid element to be unbound if it satisfies the \textit{Bernoulli criterion}, i.e., $h u_t < -1$  \citep[e.g.][]{Kastaun_2015}, where $h$ is the fluid specific enthalpy and $u_t$ is the covariant time component of the fluid element 4-velocity.\\ 

\begin{figure}
 	\includegraphics[width=\columnwidth]{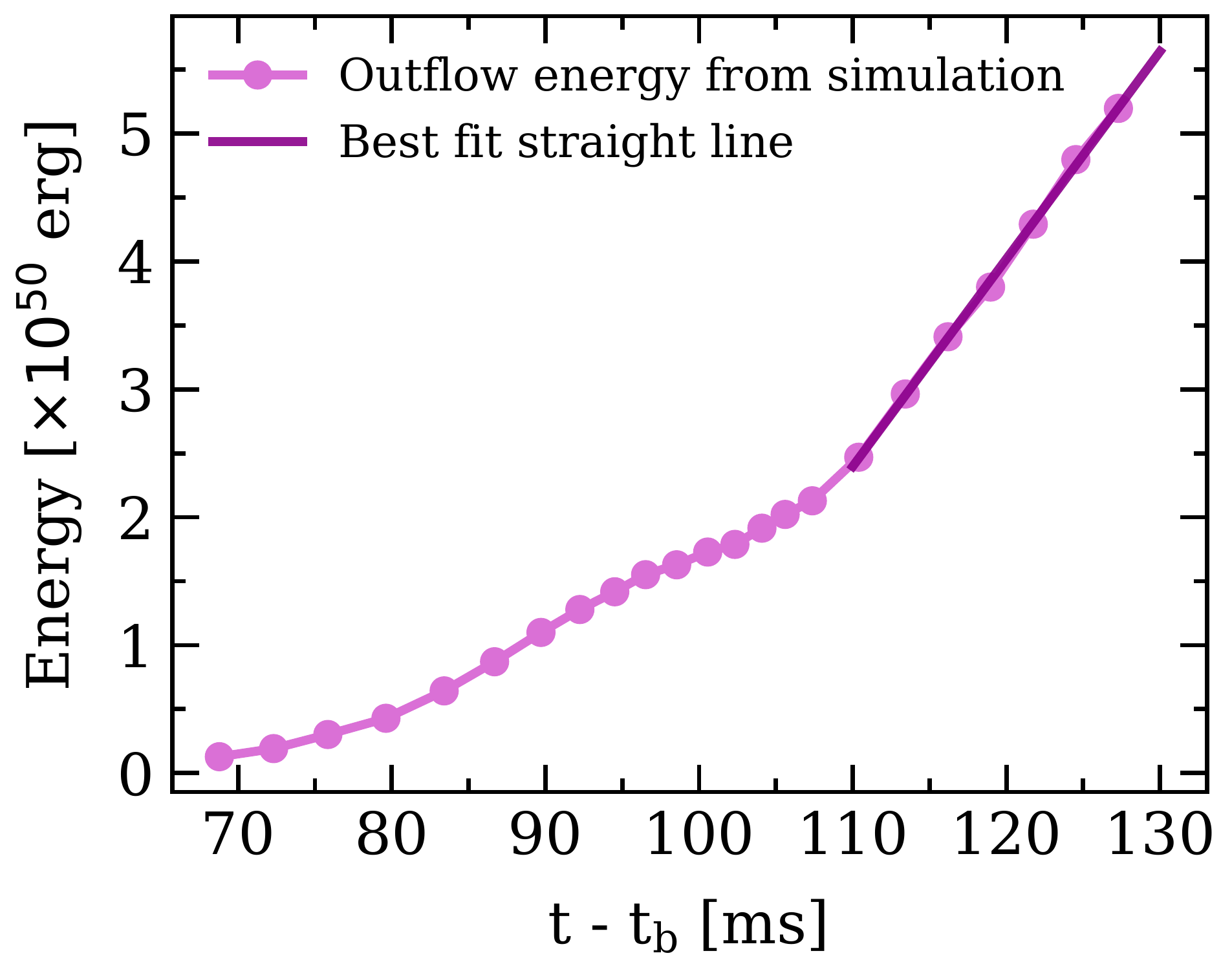}
    \caption{Total energy of jet outflow from the simulation as a function of time (total energy = kinetic energy + internal energy + magnetic energy). The outflow energy increases slowly in the beginning but shows a steady linear increase after $\sim 110$~ms. We fit the energy after $\sim 110$~ms with a straight line and use it for extrapolation in further analysis.} 
    \label{fig:energy_vs_time}
\end{figure}

We obtain the energy of the outflow as a function of time for the available data. The energy of the outflow is comprised of kinetic and internal energy of fluid elements, as well as the energy due to the magnetic field. We assume that only the unbound material forms a part of the outflow and thus do not consider bound material for the energy calculation. The time variation of the energy of outflow is shown is Fig.~\ref{fig:energy_vs_time}. We find that the energy increases slowly at first, but shows a steady linear increase after $\sim 110$~ms postbounce, the time at which an outflow along the rotation axis is launched. We use the energy evolution after this time for extrapolation. \\
         
\begin{figure}
 	\includegraphics[width=\columnwidth]{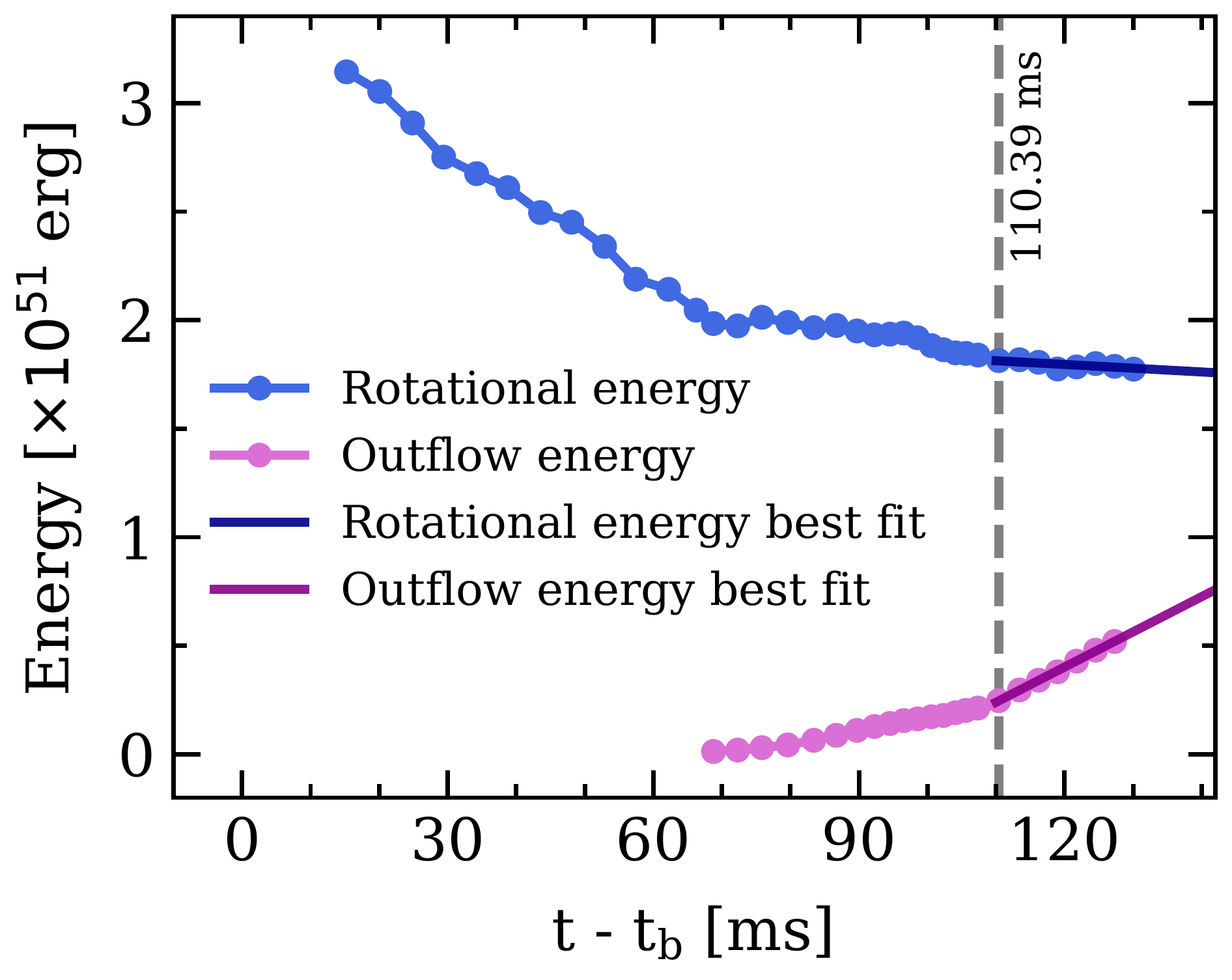}
    \caption{Rotational energy of material within a radius of $50$~km as a function of time. We also show the jet outflow energy for comparison. The decrease in rotational energy provides the energy for the outflow. We fit the rotational energy after $\sim 110$~ms with a straight line and use it for extrapolation in further analysis. } 
    \label{fig:50_km_RE_TE}
\end{figure}
         
\begin{figure}
 	\includegraphics[width=\columnwidth]{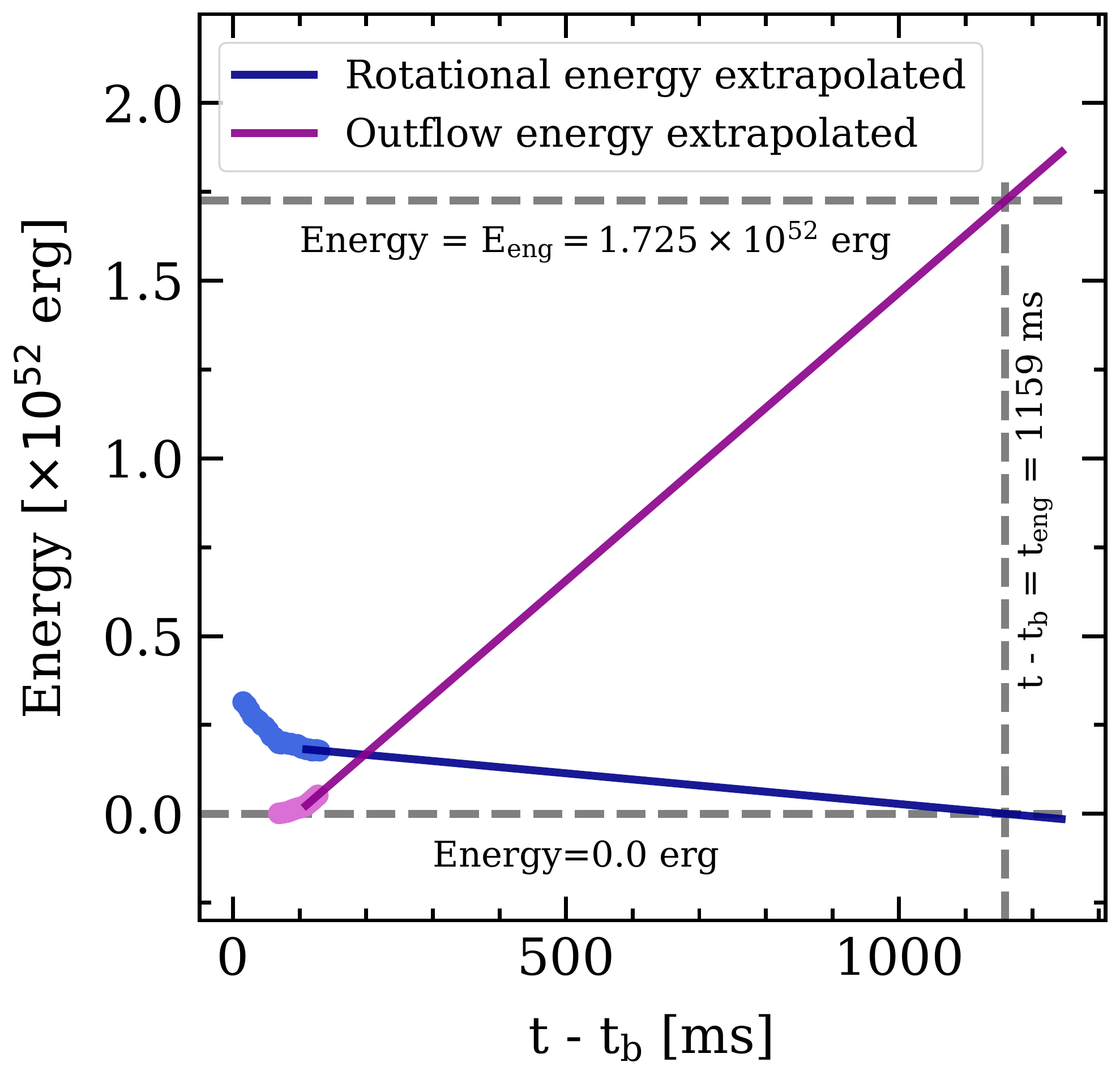}
    \caption{Extrapolation of rotational energy (within 50~km) and outflow energy to obtain the total energy, $E_{\text{eng}}$, and the characteristic time-scale, $t_{\text{eng}}$, of the central engine. We calculate $t_{\text{eng}}$ as the time when the extrapolated rotational energy becomes zero ($1159$~ms in this case). The value of the extrapolated outflow energy at this time gives $E_{\text{eng}}$ ($1.725 \times 10^{52}$~erg in this case).} 
    \label{fig:50_km_RE_TE_extrapolated}
\end{figure}

We also calculate the rotational energy of the PNS as a function of time. The rotational energy of the newly formed PNS is affected by the infalling stellar material as well as the outflow of material from the core. The infalling material imparts angular momentum and thus tends to increase the rotational energy, whereas the material in the outflow tends to decrease the rotational energy. We plot the time dependence of the rotational energy of the material within a radius of $50$~km of the centre of the PNS in Fig.~\ref{fig:50_km_RE_TE}. We find that the rotational energy overall decreases over time. Energy lost from the overall decrease in rotational energy provides the energy of the central engine. We fit the rotational energy in the 3D CCSN simulation after $\sim 110$~ms postbounce with a straight line. In order to get a limiting case, we calculate the time at which this straight line fit leads to zero rotational energy. We use this time as the characteristic time-scale ($t_{\text{eng}}$) of the central engine. We use $t_{\text{eng}}$ to extrapolate the linear part of outflow energy assuming that the outflow energy variation remains linear up to $t_{\text{eng}}$. The extrapolated outflow energy value at $t_{\text{eng}}$ gives the total energy of the central engine.  We demonstrate this method in Fig.~\ref{fig:50_km_RE_TE_extrapolated}. It should be noted that the total rotational energy initially cannot be assumed to be the total energy of the central engine because accretion of material into the vicinity of the PNS adds energy to the engine over time. \\

\begin{figure}
 	\includegraphics[width=\columnwidth]{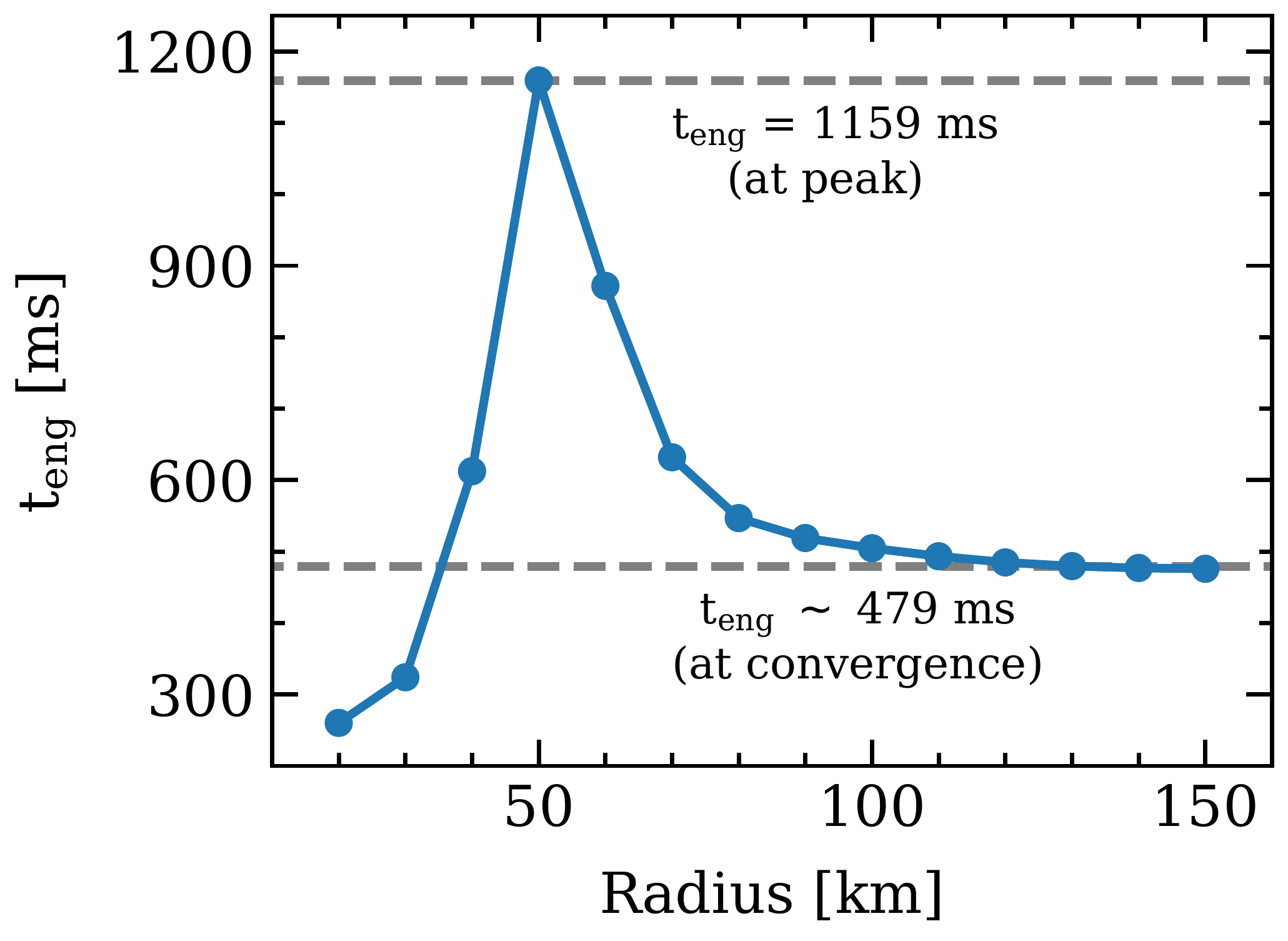}
    \caption{ $t_{\text{eng}}$ is calculated as the time at which the rotational energy of the material within a certain radius becomes zero. We show $t_{\text{eng}}$ as a function of radius in the figure. $t_{\text{eng}}$ peaks at the radius of $50$~km and converges for larger radii. We choose the value of $t_{\text{eng}}$ at peak ($1159$~ms) and convergence ($\sim 479$~ms) as the two limiting cases for further investigation.} 
    \label{fig:tjet_vs_nkm}
\end{figure}

Since we are extrapolating the rotational energy to calculate $t_{\text{eng}}$, it is important that we consider the possible parameter dependence of $t_{\text{eng}}$. To do so, we vary the radius within which material is considered for the rotational energy calculation. We vary the radius from 20~km to 150~km. For each radius, we fit the rotational energy after $\sim 110$~ms postbounce with a straight line and calculate $t_{\text{eng}}$ by extrapolating this straight line fit to zero rotational energy. We show $t_{\text{eng}}$ as a function of radius in Fig.~\ref{fig:tjet_vs_nkm}. We find that $t_{\text{eng}}$ has a peak at $\sim 1159$~ms and it converges to $\sim 479$~ms for larger radii. We therefore explore $t_{\text{eng}}=1159$~ms and $t_{\text{eng}}=479$~ms as limiting cases for this work. Extrapolating the linear part of the outflow energy (after $\sim 110$~ms postbounce) to these characteristic time-scales gives an engine energy of $1.725 \times 10^{52}$~erg for $t_{\text{eng}}=1159$~ms and $6.231 \times 10^{51}$~erg for $t_{\text{eng}}=479$~ms. We present these parameters in Table~\ref{tab:energy_table}.
         
\begin{table}
	\centering
	\caption{Extracted values of characteristic time-scale and total energy of the central engine}
	\label{tab:energy_table}
	\begin{tabular}{ccc} 
		\hline
		 Radius & Characteristic time-scale &  Total engine energy   \\
		   (km) & $t_{\text{eng}}$ (s)	 & $E_{\text{eng}}$ (erg) \\
		\hline
		   50 & 1.159   &   $1.725 \times 10^{52}$ \\
		\hline	
		  $\sim 130$  &   0.479   &   $6.231 \times 10^{51}$ \\
		\hline	
	\end{tabular}
\end{table}

\subsubsection*{Estimation of the half-opening angle}
To estimate the half-opening angle of the central engine from the wide-lobed outflow of the PNS, we use the fluid elements which are highly magnetized. We use the plasma $\beta$ parameter, defined as $\beta= P_{\text{gas}}/P_{\text{mag}}$. For highly magnetized material, $\beta \ll 1$. The jet consists of highly magnetized material and we choose two separate cases to identify the material in the jet for the calculation of the opening angle: $\beta \sim 0.1$ and $\beta \sim 0.3$. We do this because the boundary of the jet is not precisely defined and there is not a single fixed value of $\beta$ that we can use to identify the jet boundary. These two choices of $\beta$ allow us to explore a wider range of opening angles for their ability to produce SNe Ic-bl. In addition, we only consider unbound fluid elements using the \textit{Bernoulli criterion}. \\
 
\begin{figure}
 	\includegraphics[width=\columnwidth]{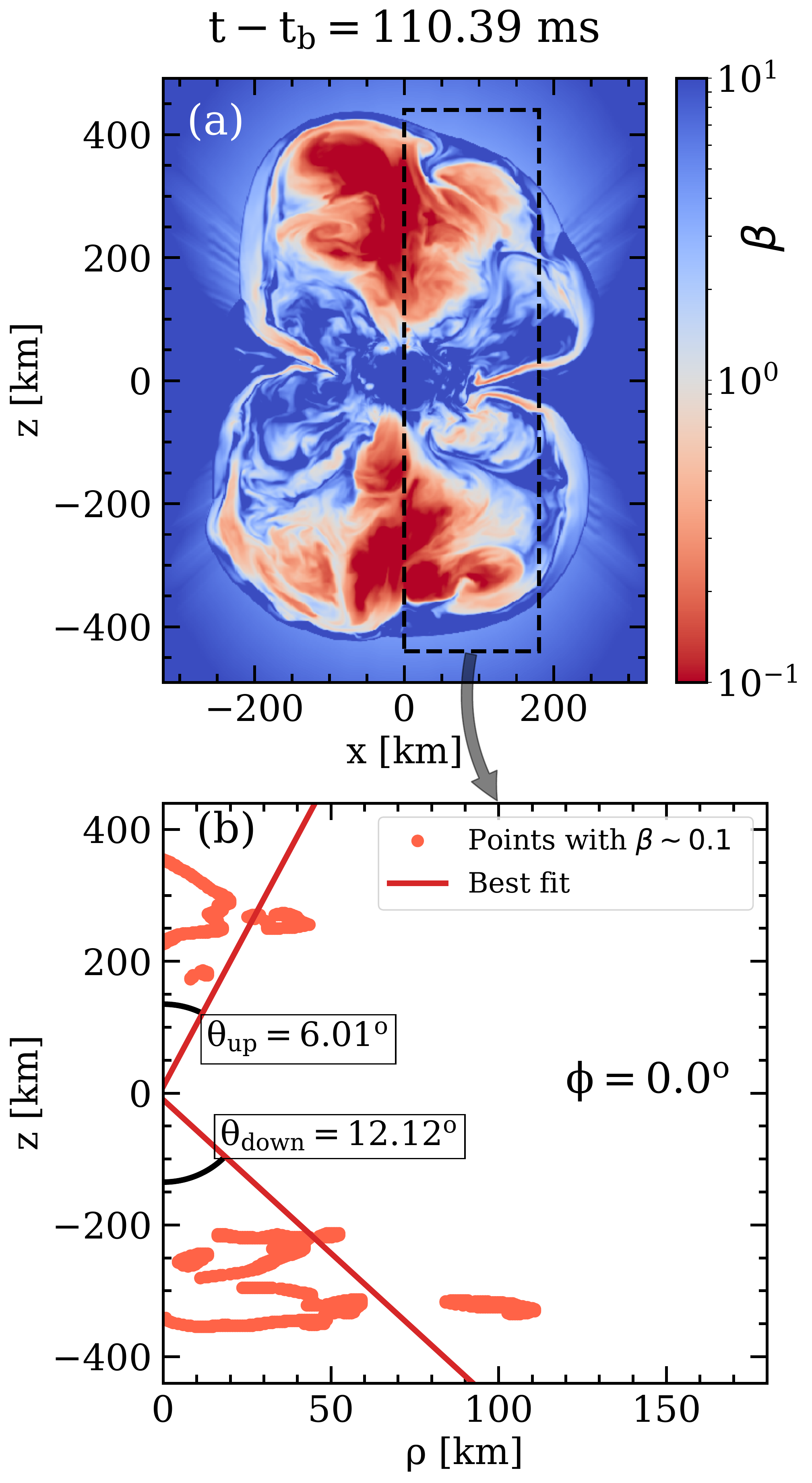}
    \caption{Panel (a) shows $\beta = P_{\text{gas}}/P_{\text{mag}}$ in the $x$-$z$ plane ($y=0$) in Cartesian coordinates at $\sim 110$~ms postbounce. The PNS is located at the origin. The 3D data in Cartesian coordinates ($x,y,z$) is converted to Cylindrical coordinates ($\rho,\phi,z$). Panel (b) shows the data in Cylindrical coordinates in the $\rho$-$z$ plane ($\phi=0$). Only those points are selected which are unbound, have positive outward $z$-velocity and have $\beta=0.1 \pm 0.005$. The selected points are fit with a straight line to obtain the effective opening angles ($\theta_{\text{up}}, \theta_{\text{down}}$) for this $\phi$-slice ($\phi = 0$). The averaging of these angles for all $\phi$-slices gives the effective opening angle $\theta_{\text{avg}}(t)$ for this timestep.} 
    \label{fig:plasmaBeta_angle_fit}
\end{figure}

Data from the 3D CCSN simulation is available in Cartesian coordinates ($x,y,z$) where $z$ is the axis of rotation of the PNS. We convert this data to cylindrical coordinates ($\rho,\phi,z$) so that averaging the obtained angles in the azimuthal direction becomes convenient. For a given timestep, at a given $\phi$-slice, we locate the fluid elements which are unbound, have positive outward $z$-velocity and have $\beta = 0.1 (0.3) \pm 0.005$. We separate these points in two regions: up ($z>0$) and down ($z<0$), and find the angles for these two regions separately. For each region, we fit the selected fluid elements with a straight line passing through the PNS surface ($\sim 15$~km). We determine the angle of this line with the z-axis as $\theta_{\text{up}}(\phi,t)$ for $z>0$  and $\theta_{\text{down}}(\phi,t)$ for $z<0$. We demonstrate the calculation of $\theta_{\text{up}}$ and $\theta_{\text{down}}$ at $t \sim 110$~ms postbounce for $\phi=0$ in Fig.~\ref{fig:plasmaBeta_angle_fit}. We average $\theta_{\text{up}}$ and $\theta_{\text{down}}$ over all values of $\phi$ to get $\theta_{\text{up,avg}}(t)$ and $\theta_{\text{down,avg}}(t)$ respectively. We calculate the average half-opening angle for the entire timestep as $\theta_{\text{avg}}(t) = \frac{1}{2}[\theta_{\text{up,avg}}(t) + \theta_{\text{down,avg}}(t)]$. This azimuthal averaging of angles minimizes the differences due to any possible small tilt of rotation axis with the $z$-axis and thus gives us an effective half-opening angle. \\

\begin{figure}
 	\includegraphics[width=\columnwidth]{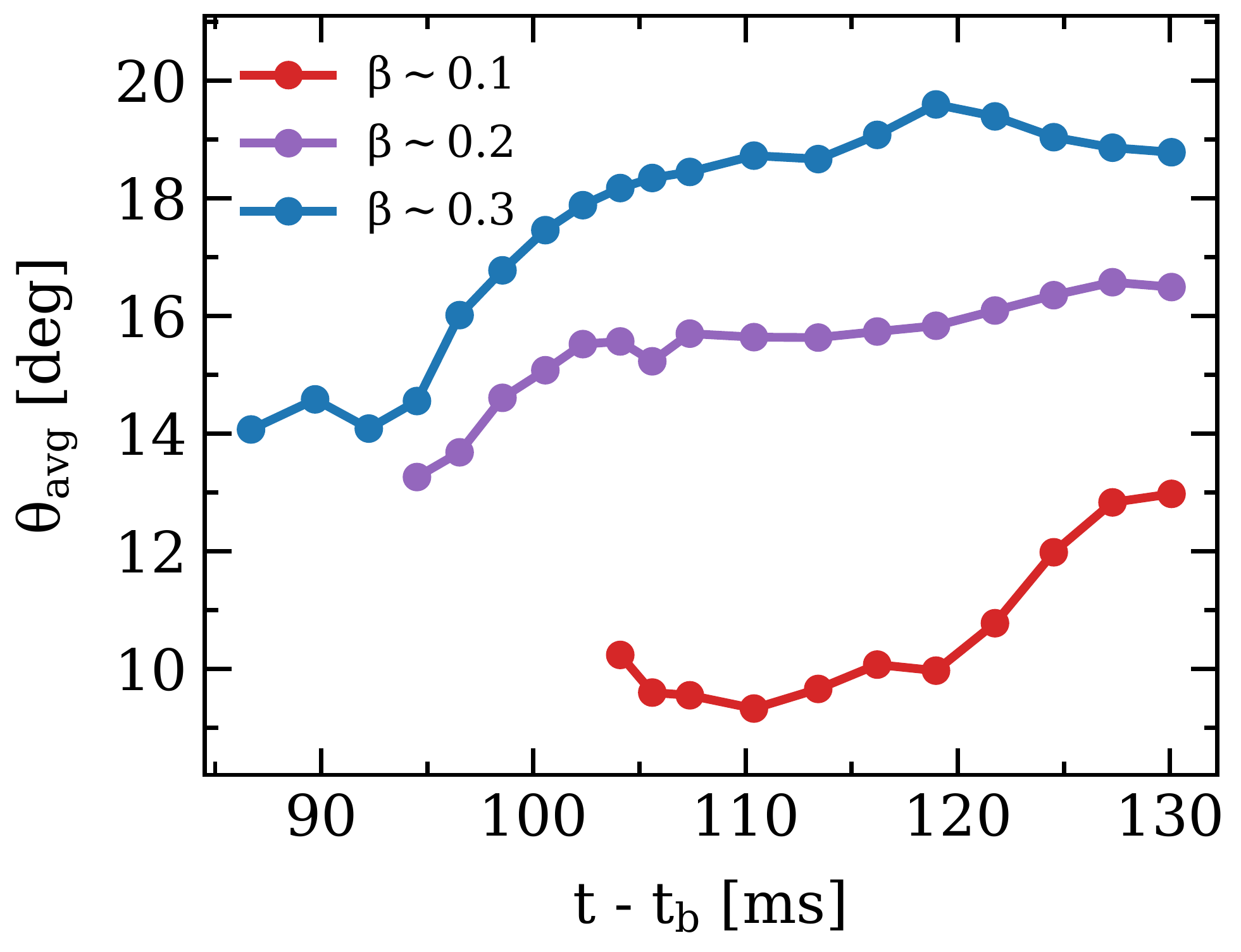}
    \caption{$\theta_{\text{avg}}$ as a function of time for $\beta$ = 0.1, 0.2 and 0.3. The opening angles become larger for higher values of $\beta$, because higher $\beta$ corresponds to less magnetized fluid elements, thus farther away from the jet axis. We consider $\beta$ = 0.1 and 0.3 for central engine parameter estimation to test dependence on $\beta$. } 
    \label{fig:angle_vs_time}
\end{figure}

\begin{figure}
 	\includegraphics[width=\columnwidth]{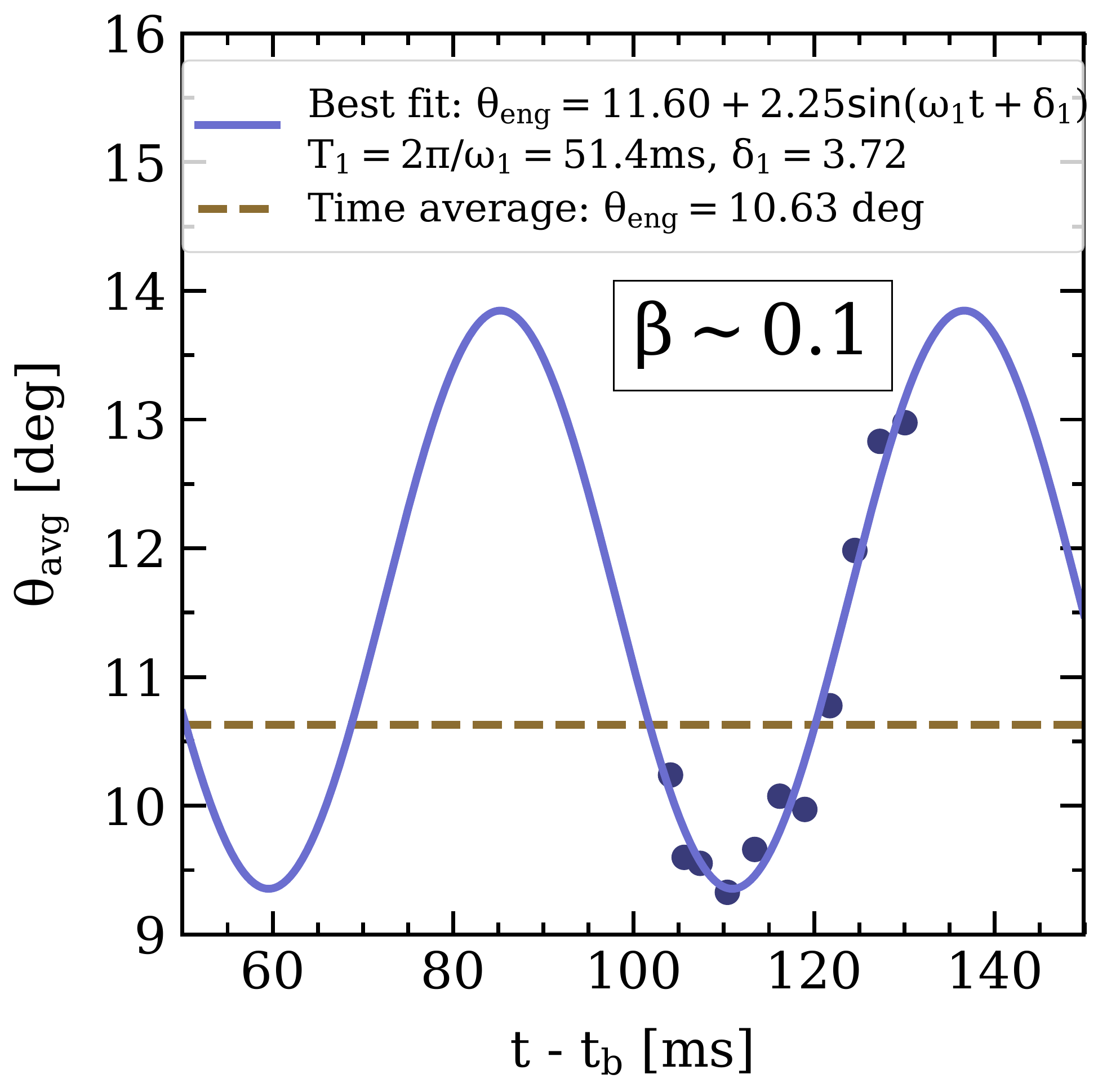}
 	\includegraphics[width=\columnwidth]{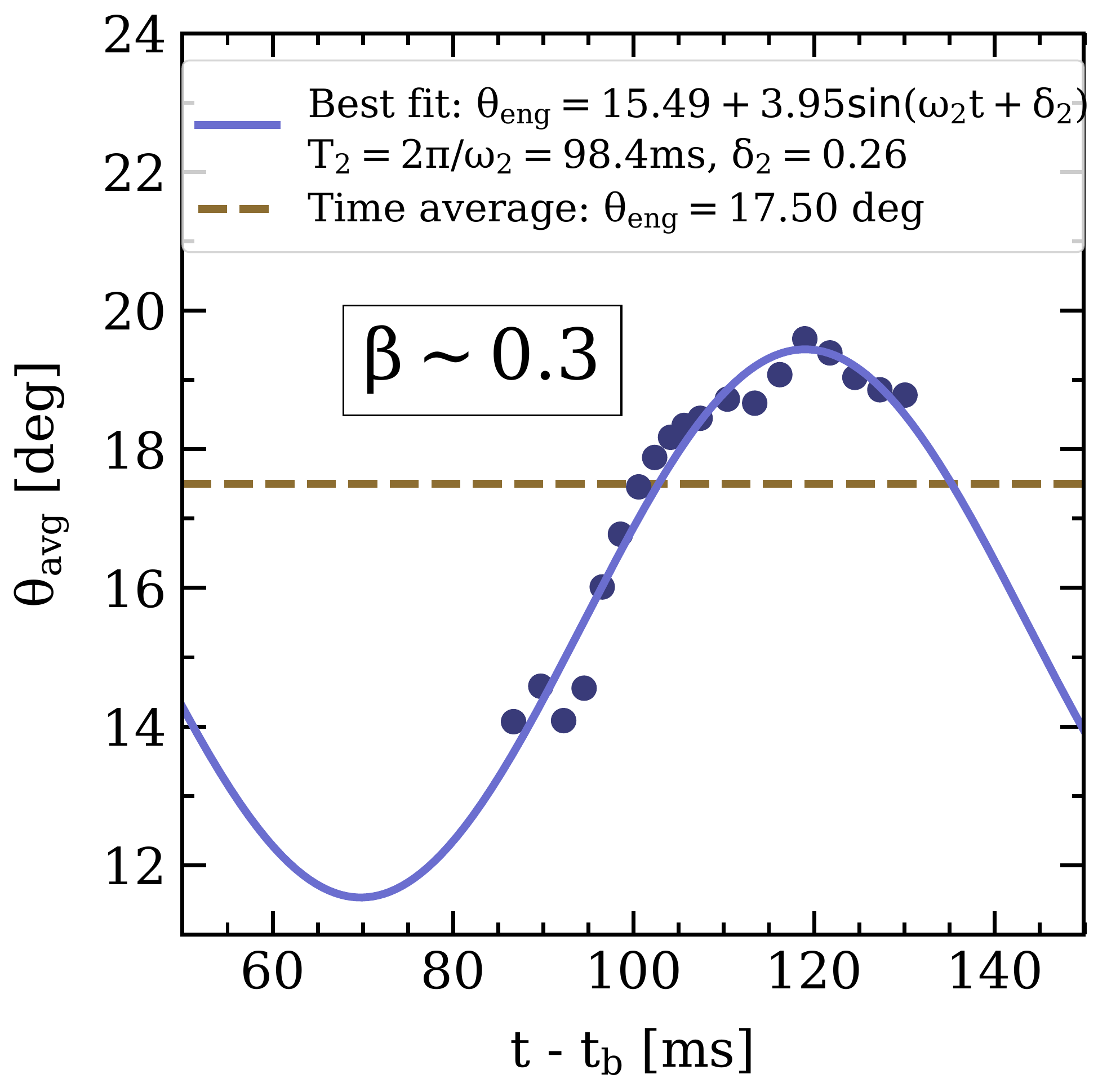}
    \caption{Extraction of $\theta_{\text{eng}}$ from $\theta_{\text{avg}}$ for $\beta$ = 0.1 and 0.3. We consider two cases for extraction of $\theta_{\text{eng}}$ for each $\beta$: (i) $\theta_{\text{eng}}$ = time average of $\theta_{\text{avg}}(t)$  (ii) $\theta_{\text{eng}}$ = sinusoidal best fit of $\theta_{\text{avg}}(t)$. Case (i) gives a constant $\theta_{\text{eng}}$, while case (ii) gives a time variable $\theta_{\text{eng}}$.} 
    \label{fig:angle_vs_time_bestfit}
\end{figure}

We calculate $\theta_{\text{avg}}$ for all values of $t$ where it is feasible, which comes out to be $105 \lesssim t-t_{\text{b}} \lesssim 130$~ms for $\beta \sim 0.1$, and $86 \lesssim t-t_b \lesssim 130$~ms for $\beta \sim 0.3$. For times earlier than $105(86)$~ms, the magnetic fields in the not yet fully formed outflow are not strong enough to produce $\beta$ as low as $0.1(0.3)$ for all $\phi$-slices. We show the time dependence of $\theta_{\text{avg}}$ for $\beta=0.1$, $0.2$ and $0.3$ in Fig.~\ref{fig:angle_vs_time}. We perform the remaining analysis only for $\beta=0.1$ and $\beta=0.3$ as we are interested in the limiting values of the possible opening angle. We find that $\theta_{\text{avg}}$ varies from $\sim 9(14)\degr$ to $\sim 13(19)\degr$ for $\beta=0.1(0.3)$. In order to get the half-opening angle of the central engine from this data, we take 2 cases. In the first case we take the average over time for $\theta_{\text{avg}}(t)$. As the outflow in the 3D simulation effectively precesses in time due to the kink instability, we try to parametrize this via a sinusoidal varying in time. In this second case we fit $\theta_{\text{avg}}$ as  $\theta(t) = A \sin(\omega t + \delta) + B$. We show the data points as well as the best fit in Fig.~\ref{fig:angle_vs_time_bestfit}. We summarize the extracted opening angles for all cases in Table~\ref{tab:angle_table}. 
	
		\begin{table}
	\centering
	\caption{Extracted values of opening angles of the central engine }
	\label{tab:angle_table}
	\begin{tabular}{ccc} 
		\hline
		$\beta$ & Time averaged $\theta_{\text{avg}}(t)$: & Sinusoidal fit of $\theta_{\text{avg}}(t)$:   \\
		 &  $\theta_{\text{eng}}$(deg) & $\theta_{\text{eng}}(t)$(deg) \\
		\hline
		0.1 & 10.63 & 11.60+2.25$\sin(\omega_1 t+\delta_1)$  \\
		    &       & $T_1=\frac{2\pi}{\omega_1} =51.4$~ms, $\delta_1=3.72$  \\
		\hline
		0.3 & 17.50 & 15.49+3.95$\sin(\omega_2 t+\delta_2)$  \\
		    &       & $T_2=\frac{2\pi}{\omega_2}=98.4$~ms, $\delta_2=0.26$  \\
		\hline
	\end{tabular}
\end{table}

\subsubsection*{Summary of parametric central engine models}
We have extracted two different values of ($E_{\text{eng}},t_{\text{eng}}$), as summarized in Table~\ref{tab:energy_table}. For $\theta_{\text{eng}}$, we extracted four different values: two constant in time and two varying in time, as summarized in Table~\ref{tab:angle_table}. We combine these parameters and construct eight parametric models for the central engine. They are summarized in Table~\ref{tab:run_table}. Model 1 is the reproduction of the work of \textit{B2018}. We find that the extracted parameters in model 2 are very close to the values used by \textit{B2018}. Model 2 consists of ($E_{\text{eng}}, t_{\text{eng}}$) extracted from peak $t_{\text{eng}}$, and $\theta_{\text{eng}}$ extracted from time averaged $\theta_{\text{avg}}(t)$ for $\beta \sim 0.1$. We perform hydrodynamic and radiation-transfer calculations for models 2 to 8, and determine whether they are able to produce a SN Ic-bl. Among these models we judge model 2 to be most realistic because (i) $\beta \sim 0.1$ identifies the most highly magnetized jet particles and (ii) $t_{\text{eng}} = 1179$~ms is the time-scale extracted at $r=50$~km from the centre of PNS. This is approximately in the middle of possible PNS radii which vary from $30$~km to $80$~km \citep{Glas_2019}. It also allows us to explore the largest extent of parameter space  because $t_{\text{eng}}$ has a peak at $50$~km. 

	\begin{table}
	\centering
	\caption{Central engine models constructed from the extracted parameters presented in Table~\ref{tab:energy_table} and Table~\ref{tab:angle_table} }
	\label{tab:run_table}
	\setlength\tabcolsep{4pt} 
	\begin{tabular}{cccc}
		\hline
		Engine Model & $t_{\text{eng}}$(s) & $E_{\text{eng}}$(erg) & $\theta_{\text{eng}}$(deg)   \\
		\hline
		Model 1 & 1.1 & $1.8 \times 10^{52}$ & 11.5  \\
		\hline
		Model 2 & 1.159 & $1.725 \times 10^{52}$ & 10.63   \\
		\hline
		Model 3 & 1.159 & $1.725 \times 10^{52}$ & 17.50  \\
		\hline
		Model 4 & 0.479 & $6.231 \times 10^{51}$ & 10.63  \\
		\hline
		Model 5 & 0.479 & $6.231 \times 10^{51}$ & 17.50   \\
		\hline
		Model 6 & 1.159 & $1.725 \times 10^{52}$ & $11.60 + 2.25 \sin(\omega_1 t + \delta_1)$ \\
		\hline
		Model 7 & 1.159 & $1.725 \times 10^{52}$ & $15.49 + 3.95 \sin(\omega_2 t + \delta_2)$   \\
		\hline
		Model 8 & 0.479 & $6.231 \times 10^{51}$ & $11.60 + 2.25 \sin(\omega_1 t + \delta_1)$  \\
		\hline
		Model 9 & 0.479 & $6.231 \times 10^{51}$ & $15.49 + 3.95 \sin(\omega_2 t + \delta_2)$   \\
		\hline
	\end{tabular}
\end{table}

\subsection{Supernova Observables}
\begin{figure*}
 	\includegraphics[width=0.95\textwidth]{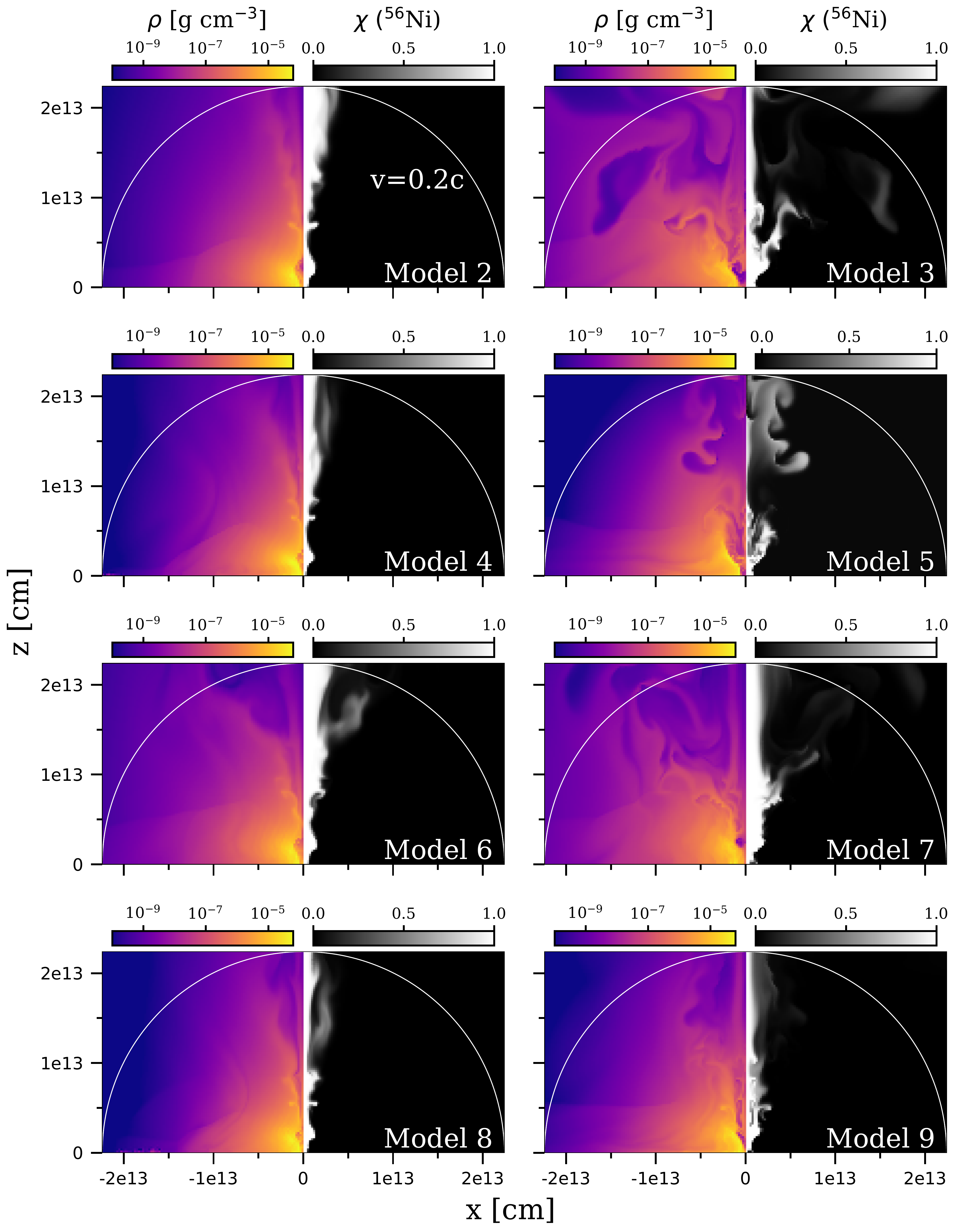}
    \caption{ Mass density, $\rho$ (left panel) and $^{56}$Ni mass fraction, $\chi$ (right panel) at the end of the {\scriptsize{JET}} simulations ($t = 3733$~s) for models 2 to 9.  We use these snapshots as the starting point for the {\scriptsize{SEDONA}} calculations. The white line shows $v=0.2c$. } 
    \label{fig:rho_ni_distribution}
\end{figure*}

Using the central engine parameters described in Table~\ref{tab:run_table}, we perform hydrodynamic calculations using the {\scriptsize{JET}} code up to $t \sim 3700$~s. At this point, the flow becomes homologous and thus the outward velocity is proportional to the radius. We show the mass density (left panel) and the $^{56}$Ni mass fraction (right panel) at this time for models 2 to 9 in Fig.~\ref{fig:rho_ni_distribution}. At this time, the most relativistic material has reached a radius of $\sim 5\times10^{13}-1\times10^{14}$ cm depending on the model. The SN ejecta is dominated by lower velocity material ($v\lesssim 0.2c$) which extends to $\sim 2\times10^{13}$ cm. We show this region in Fig.~\ref{fig:rho_ni_distribution} and use it as the starting point for the {\scriptsize{SEDONA}} calculations. We find that the ejecta density structure shows some deviation from spherical symmetry in the form of lower-density material in an approximately conical shape around the z-axis. For models 2, 4, 6, 8 and 9, the deviations from spherical symmetry are minor, with the angle of the cone $\lesssim 10\degr$. Models 3, 5 and 7 show more deviation from spherical symmetry, with the angle of the cone $\lesssim 25\degr$. More asymmetrical ejecta in models 3, 5 and 7 is due to the higher opening angle in these models ($\sim 17\degr$). Model 9, despite having a higher opening angle, shows this behaviour to a lesser extent. The distribution of $^{56}$Ni shows much more anisotropy, with most of the $^{56}$Ni concentrated along the $z$-axis. \\

We then perform the radiation transport calculations using {\scriptsize{SEDONA}} (starting at $\sim 3700$~s) in 2D cylindrical coordinates  for the material within the region $v_{\rho}, |v_z| \leq 0.2c$. We perform the {\scriptsize{SEDONA}} simulations using 9 evenly spaced bins in $\mu = \cos(\theta)$, $\mu \in [-1,1] $, where $\theta$ is the viewing angle with respect to the polar direction. The light curves and spectra that we show are averages within the bins. \\

\begin{figure}
 	\includegraphics[width=\columnwidth]{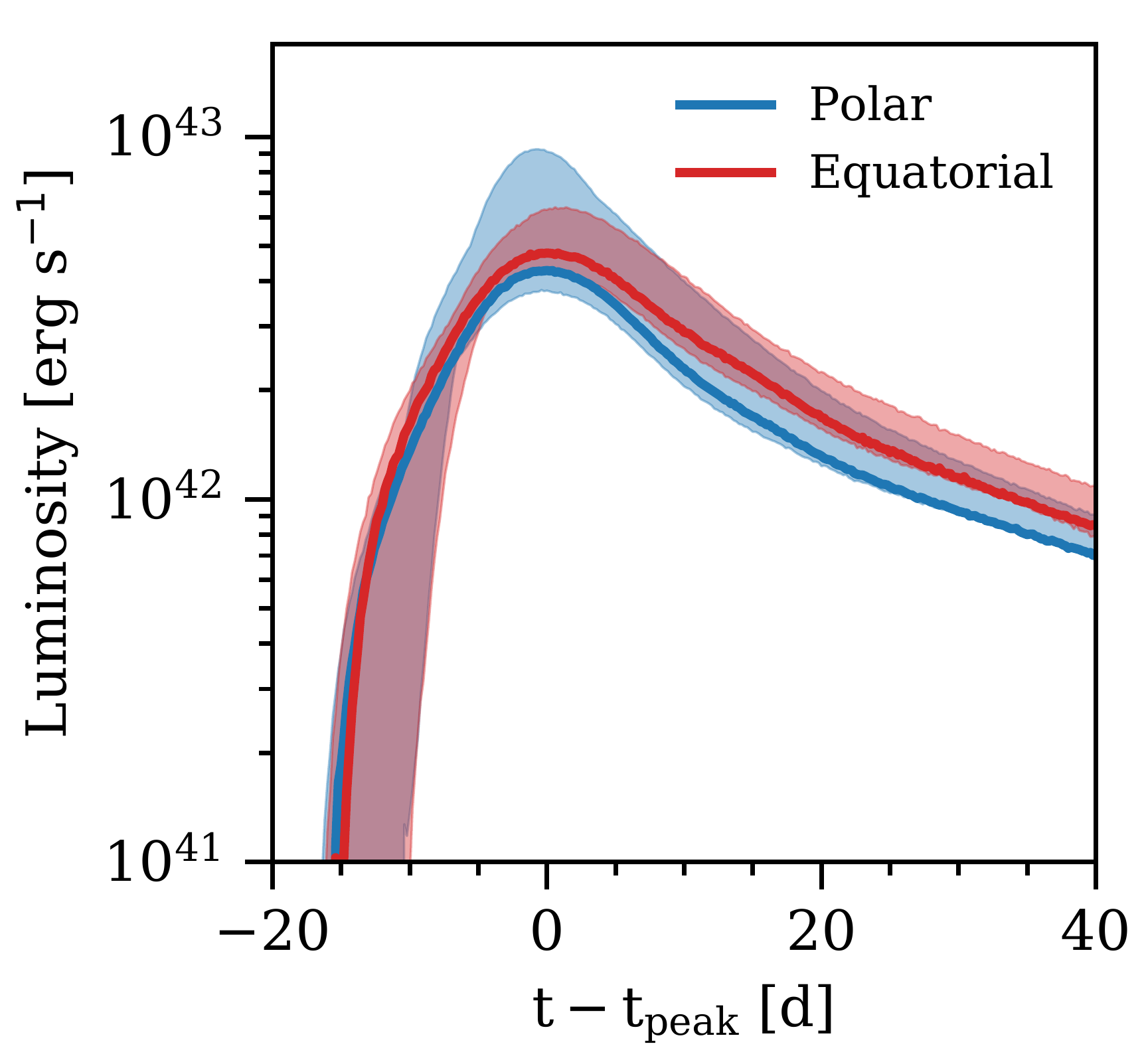}
    \caption{Bolometric light curves for our most realistic model (model 2, solid lines) along with the associated uncertainties from other models (shaded region), for polar and equatorial viewing angles. The rise times of the light curves are in the range of $\sim 10 - 18$~d, while the peak bolometric luminosities lie in the range of $\sim 4\times 10^{42} - 9\times 10^{42}$~erg s$^{-1}$. } 
    \label{fig:best_lc}
\end{figure}

\begin{table*}
	\centering
	\caption{SN and GRB properties of the various central engine models }
	\label{tab:SN_properties_table}
	\begin{tabular}{ccccccc} % four columns, alignment for each
		\hline
		Engine model & $t_\text{peak}$(d) & $L_\text{peak}$(erg s$^{-1}$) & $M(^{56}$Ni)/$M_\odot$ & Kinetic Energy (erg) & Energy in $\gamma h \gtrsim 10$ (erg) & GRB present?  \\
		 & (Eq, Po) & (Eq, Po) &  &  & &\\
		\hline
		Model 1 & (15.3, 15.1) & (5.64, 4.92)$\times 10^{42}$ & 0.21 & $4.72\times 10^{51}$ & $3.60 \times 10^{51}$ & Yes \\
		\hline
		Model 2 & (15.9, 15.9) & (4.78, 4.28)$\times 10^{42}$ & 0.18 & $3.95\times 10^{51}$ & $4.08 \times 10^{51}$ & Yes \\
		\hline
		Model 3 & (11.7, 10.1) & (6.16, 9.24)$\times 10^{42}$ & 0.24 & $9.58\times 10^{51}$ & 0.0 & No \\
		\hline
		Model 4 & (17.7, 16.9) & (4.27, 3.78)$\times 10^{42}$ & 0.16 & $3.72\times 10^{51}$ & $2.59 \times 10^{51}$ & Yes \\
		\hline
		Model 5 & (12.3, 11.1) & (6.40, 8.20)$\times 10^{42}$ & 0.24  & $5.57\times 10^{51}$ & 0.0 & No \\
		\hline
		Model 6 & (16.3, 15.3) & (5.39, 4.73)$\times 10^{42}$ & 0.20  &  $5.09\times 10^{51}$ & $1.21 \times 10^{50}$ & Yes \\
		\hline
		Model 7 & (14.1, 12.1) & (6.34, 6.93)$\times 10^{42}$ & 0.24 &  $7.34\times 10^{51}$ & $4.60 \times 10^{46}$ & No \\
		\hline
		Model 8 & (17.5, 16.7) & (4.82, 4.09)$\times 10^{42}$ & 0.18 & $3.53\times 10^{51}$ & $2.45 \times 10^{50}$ & Yes \\
		\hline
		Model 9 & (15.5, 14.5) & (6.09, 5.76)$\times 10^{42}$ & 0.23 & $5.17\times 10^{51}$ & 0.0 & No \\
		\hline
	\end{tabular}
\end{table*}

As described in the previous section, model 2 is our most realistic central engine model extracted from the CCSN simulation. We explore models 3 to 9 to account for uncertainties around our most realistic model. We show the resulting light-curves with respective uncertainties for polar and equatorial viewing directions in Fig.~\ref{fig:best_lc}, where solid lines indicate our most realistic model (model 2) and the shaded region indicates the uncertainties from the other models. We tabulate the properties of the associated SNe for all models in Table~\ref{tab:SN_properties_table}.\\

\begin{figure}
 	\includegraphics[width=\columnwidth]{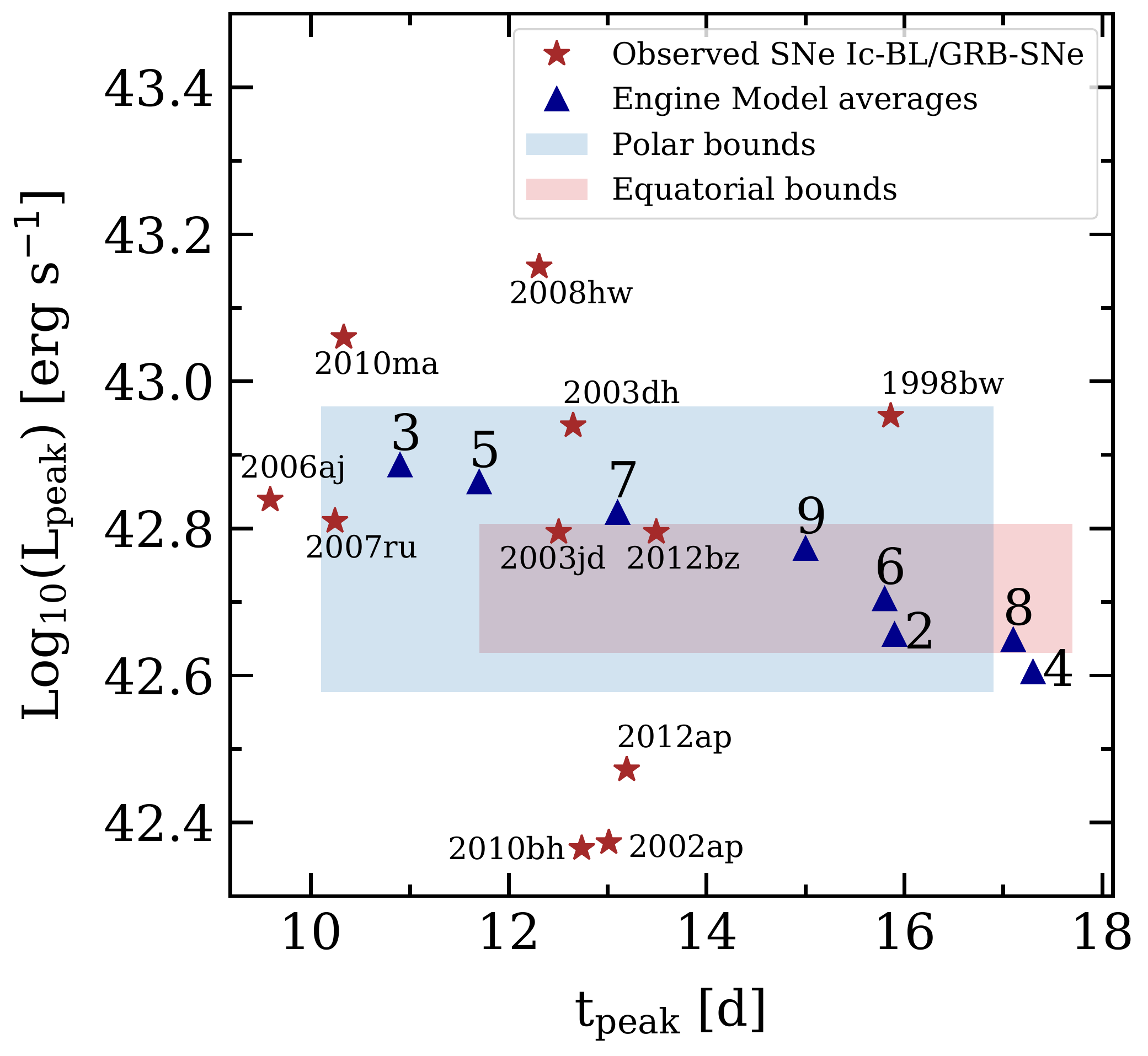}
    \caption{ Peak luminosities ($L_\text{peak}$) and rise times ($t_\text{peak}$) of light-curves of various SNe Ic-bl/GRB-SNe reconstructed by \citet{Prentice_2016} compared with the light curves of our models. The blue and red shaded regions show the bounds on the parameter space constrained by polar and equatorial light curves respectively. The blue triangles show the angle average of polar and equatorial values of ($L_\text{peak}$, $t_\text{peak}$) for our different models. Our rise times are nearly consistent with observed SNe Ic-bl, while our peak luminosities span a subset of $L_\text{peak}$'s of observed SNe Ic-bl, but are well within the observational constraints. } 
    \label{fig:Lpeak_vs_tpeak}
\end{figure}

We find that the rise times of the bolometric light curves vary between $\sim 10$ and $18$~d, while the peak luminosities vary between $\sim 4\times 10^{42}$ and $ 9\times 10^{42}$~erg~s$^{-1}$. Polar viewing angles show higher variation in peak luminosity (more than twice) compared to equatorial viewing angles. The viewing angle effect is small ($\lesssim2$~d) for the  rise times of the different models. \cite{Prentice_2016} reconstruct the pseudo-bolometric light-curves of 85 stripped-envelope SNe from the available literature, out of which 22 belong to SNe Ic-bl/GRB-SNe category. We compare $L_\text{peak}$ and $t_\text{peak}$ of their properly constrained SNe Ic-bl/GRB-SNe with the results of our models. We show the comparison in Fig.~\ref{fig:Lpeak_vs_tpeak}. The shaded regions show the parameter space spanned by our models 2 to 9 for polar and equatorial viewing angles. We see that our results span a subset of the parameter space occupied by the reconstructed SNe, with the range of rise times nearly consistent with the reconstructed SNe and the range of peak bolometric luminosities falling short of the brightest reconstructed SNe. However, all values fall well within the limits of observed SNe Ic-bl. Interestingly, we find that our models with opening angles extracted from $\beta=0.3$ (and not 0.1) show better agreement with the observed SNe Ic-bl indicating that wider outflows may fit SNe observations more easily.\\

\begin{figure}
 	\includegraphics[width=\columnwidth]{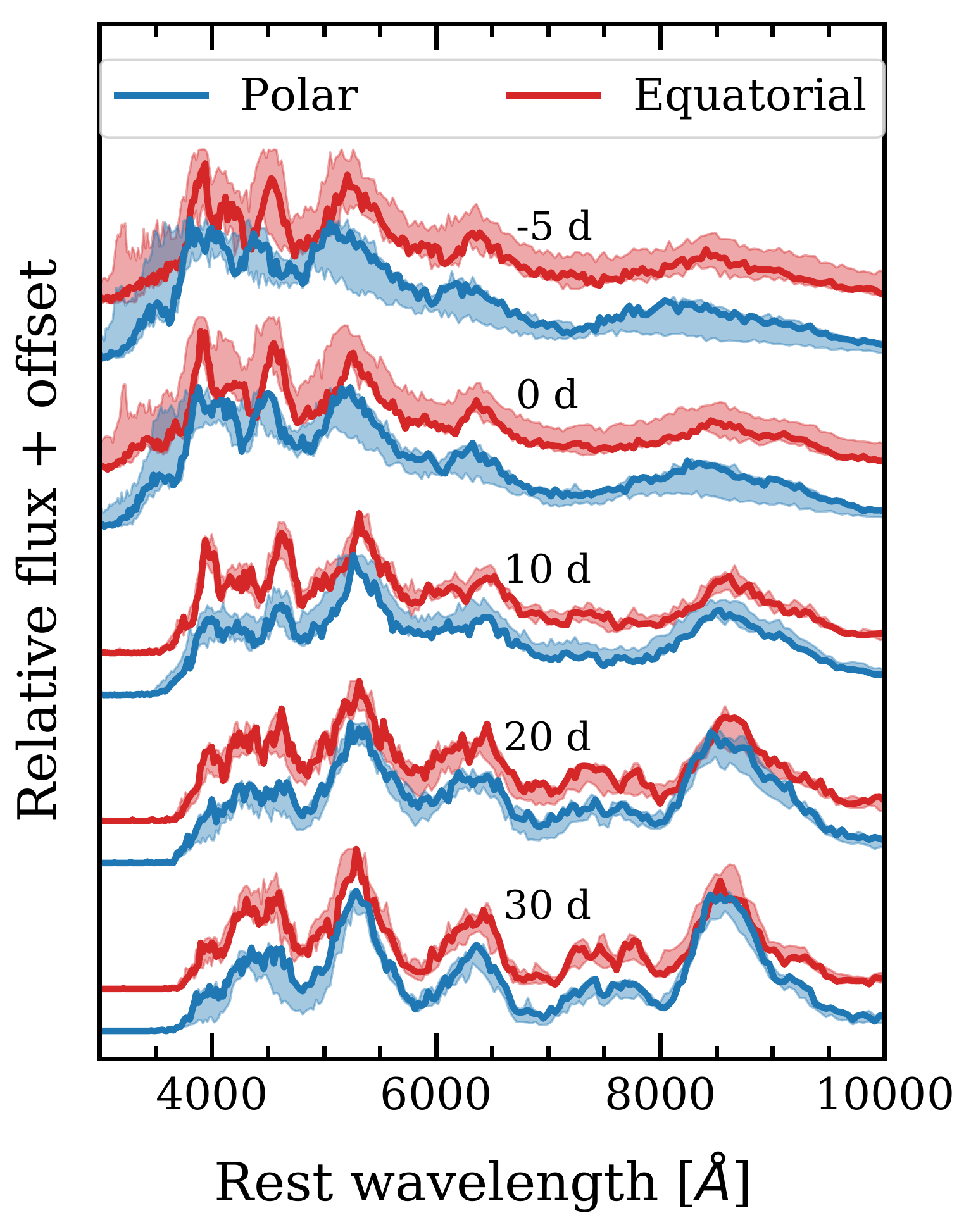}
    \caption{Time evolution of the spectra of our most realistic model (model 2, solid lines) along with the associated uncertainties from models 3 to 9 (shaded region), for polar and equatorial viewing angles. Times are relative to peak bolometric luminosity. The spectra for equatorial viewing angles have been shifted upwards with respect to polar viewing angles for compactness.  } 
    \label{fig:best_spectra}
\end{figure}

\begin{figure*}
 	\includegraphics[width=0.95\textwidth]{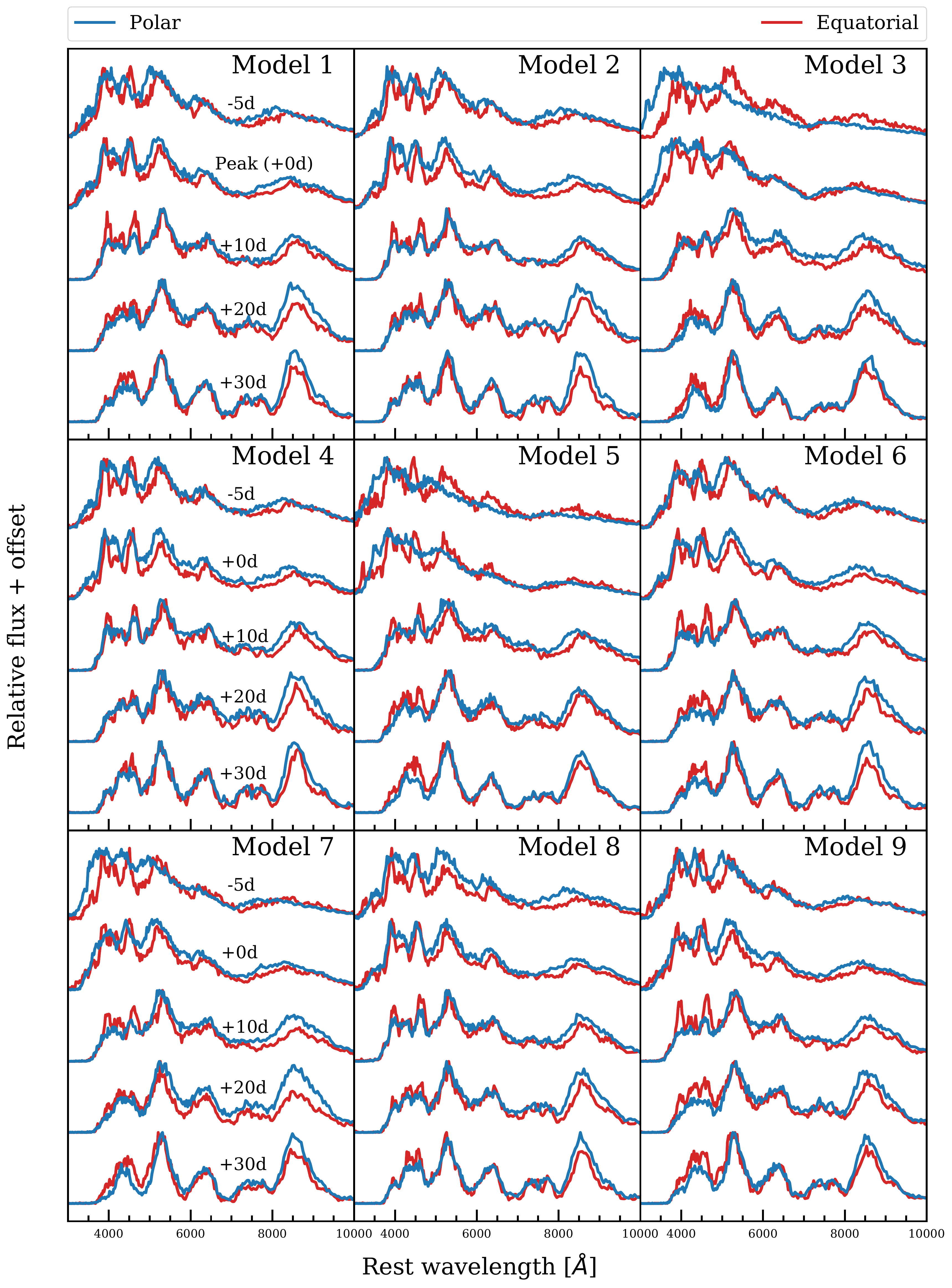}
    \caption{Time evolution of the spectra for all models for polar and equatorial viewing angles. Model 1 is the reproduction of the spectra obtained by \textit{B2018}. Model 2 is our most realistic for the 3D CCSNe simulation parameters. The differences from models 3 to 9 are the source of the shaded region in Fig.~\ref{fig:best_spectra}.  } 
    \label{fig:result_all_spectra}
\end{figure*}

We show the spectra at various times for our most realistic model (model 2) along with the associated uncertainties for polar and equatorial viewing angles in Fig.~\ref{fig:best_spectra}. The equatorial spectra have been shifted vertically with respect to the polar spectra for clarity. We show the spectra for individual models in Fig.~\ref{fig:result_all_spectra}. The viewing angle dependence is more pronounced at earlier times before the bolometric peak, while the late time spectra show little viewing angle dependence. The uncertainties are higher for both polar and equatorial viewing angles before the bolometric peak. We find that the spectra after peak show little uncertainty and thus are robust across all models. There is some model variation for late time polar spectra in the wavelength range 4000-5000~\AA, but these are due to the variation of amounts of explosively synthesized $^{56}$Ni/Co/Fe along the pole for different models, which has strong bound-bound transitions in this wavelength range and increases the line opacity differently in different models. The variation in $^{56}$Ni distribution for different models can be seen in Fig.~\ref{fig:rho_ni_distribution}.\\ 

\textit{B2018} compared their model spectra with observed SNe Ic-bl and found that their model spectra reproduce the major characteristics of SNe Ic-bl spectra. We find that our model spectra are similar to the model spectra of \textit{B2018}, with the presence of characteristic broad lines typical of a SN Ic-bl. Also, our spectra for $t>t_\text{peak}$ are nearly consistent across various models and viewing angles. We conclude that within this end-to-end simulation setup, the jetted outflow from a rapidly rotating 3D CCSN simulation is compatible with SNe Ic-bl light curves and spectra. \\

A SN Ic-bl can be launched even if the jet engine fails to produce a GRB. We use the scaled terminal Lorentz factor $\gamma h$, where $\gamma$ is the Lorentz factor and $h$ is the specific enthalpy of the fluid scaled by $c^{-2}$, to determine whether a particular engine model produces a GRB. We track the evolution of $\gamma h$ and assume that material with $\gamma h \gtrsim 10$ post breakout constitutes a GRB if the total energy in the material with $\gamma h \gtrsim 10$ is greater than $\sim 10^{50}$~erg. We find that models 1, 2, 4, 6 and 8, which have $\theta_{\text{eng}} \sim 10\degr$, produce a successful GRB. We do not observe a GRB in models 3, 5, 7 and 9, which have $\theta_{\text{eng}} \sim 16\degr$. This implies that narrower jet outflows provide more suitable conditions for the formation of GRBs. SN Ic-bl's without GRBs may be associated with wider outflows. Analyzing the GRB properties in more detail will be carried out in future work.  

\section{Conclusions and Discussion}\label{discussion}
\textit{B2018} carried out end-to-end SRHD and radiation transfer simulations with a single central engine and successfully produced light curves and spectra nearly consistent with observations of SNe Ic-bl for a presumed set of engine parameters. In this work we have extended their numerical setup and instead of presuming the values of engine parameters we extract them from the 3D CCSN simulations of \cite{Moesta_2014}. We find that the range of light curves obtained in our setup have nearly consistent rise times with observations of SNe Ic-bl and their peak bolometric luminosities ($L_\text{peak}$) form a subset of the full range of  $L_\text{peak}$ observed in SNe Ic-bl. We also find that our spectra for $t>t_\text{peak}$ are fairly robust across our different models and viewing angles, and are similar to the spectra of \textit{B2018}. Due to this similarity and the presence of characteristic broad spectral features we conclude that our spectra are consistent with the spectra of SNe Ic-bl. This indicates that jet outflows produced in rapidly-rotating CCSNe explosions can successfully trigger a SN Ic-bl.\\

There is uncertainty regarding the most accurate method of parameter extraction from the 3D CCSNe simulations. To account for that we extract a range of possible values for the parameters and investigate the uncertainties arising from these in the light curves and spectra. We use the effective rate of decrease of rotational energy of the PNS to determine the engine duration and energy. The radius within which we calculate the rotational energy is not a single precise value, and to explore this uncertainty we have used a range of possible values. Another uncertainty is our assumption that the rate of decrease of rotational energy is constant for the entire extrapolated time, which is nearly an order of magnitude larger ($\sim 1000$~ms) compared to the available data ($\sim 100$~ms). In reality, the rate of change of rotational energy will depend on the dynamic interplay between the jet outflow and the infalling stellar material. Similarly, we have used the plasma $\beta$ parameter, which is lower for highly magnetized material, to determine the opening angle of the jet. We know that material within the jet has very low $\beta$, but there isn't a single precise value that determines the boundary of the jet. To account for that, we choose $\beta$ between 0.1 and 0.3 and assume that these values approximate the jet boundary reasonably well.\\

Our most realistic $E_{\text{eng}}$ and $t_{\text{eng}}$ comes from the rate of decrease of the rotational energy for material within a \textit{50~km} radius. Our most realistic $\theta_{\text{eng}}$ comes from the material with $\beta \sim 0.1$. This provides our most realistic model (model 2) with parameters $E_{\text{eng}} = 1.725\times10^{52}$~erg, $t_{\text{eng}} = 1.159$~s and $\theta_{\text{eng}} = 10.63\degr$. Interestingly, this is very close to the engine parameters presumed by \textit{B2018} ($E_{\text{eng}} = 1.8\times10^{52}$~erg , $t_{\text{eng}} = 1.1$~s and $\theta_{\text{eng}} = 11.5\degr$). However, in the analysis of our other models, we find that our models which use $\theta_{\text{eng}}$ extracted from the material with $\beta \sim 0.3$ (extracted $\theta_{\text{eng}}\sim 17\degr$) show better agreement with observed $t_\text{peak}$ and $L_\text{peak}$ of SN Ic-bl light curves compared to the models using $\beta \sim 0.1$ (extracted $\theta_{\text{eng}}\sim 11\degr$). This is likely due to the jet coupling more easily to the stellar material for larger opening angles which also leads to a higher $^{56}$Ni mass synthesized during the explosion.\\
 
The {\scriptsize{JET}} simulations performed in this work are 2D. In future work we plan to use 3D {\scriptsize{JET}} and {\scriptsize{SEDONA}} simulations and include magnetic fields. We have not included magnetic fields in this work because MHD simulations in 2D vs 3D show fundamentally different results \citep{Moesta_2014,Bromberg_2016}. We will also explore more accurate methods for extracting central engine parameters from the CCSN simulation data in future work. There is also a need for longer 3D CCSN simulation data to better extract the late time behaviour of the engine. In this way we can reduce the uncertainty arising from the extrapolation over a time-scale of $1$~s. This is challenging due to high computational cost of 3D CCSN simulations ($\sim$ a month on $\sim 1000$ nodes) but will be enabled by the advent of GPU-based codes which will reduce the computational time  considerably and lead to availability of data over longer time scales. This work uses the central engine parameters extracted from a single CCSN simulation, which may not be  representative of the full extent of possible parameters. Performing the present analysis for  other CCSN simulations will help check the consistency of the current results as well as explore the full range of possible parameters, and then the  full range of possible light curves and spectra.  
 
\section*{Acknowledgements}\label{ack} 
  
The authors would like to thank A.~Tchekhovskoy for discussions. The simulations were carried out on NCSA's
BlueWaters under NSF awards PRAC OAC-1811352 (allocation PRAC\_bayq), NSF AST-1516150 (allocation PRAC\_bayh), and allocation ILL\_baws, and TACC's Frontera under allocation DD FTA-Moesta. We thank SURFsara (\url{www.surfsara.nl}) for the support in using the Lisa Compute Cluster.

%%%%%%%%%%%%%%%%%%%%%%%%%%%%%%%%%%%%%%%%%%%%%%%%%%



%\bibliographystyle{mnras}
\bibliographystyle{mnras}
\bibliography{references}
%\bibliographystyle{ieeetr}

% Don't change these lines
\bsp	% typesetting comment
\label{lastpage}
\end{document}